\documentclass[letterpaper]{article} 
\usepackage{aaai25}  
\usepackage{times}  
\usepackage{helvet}  
\usepackage{courier}  
\usepackage[hyphens]{url}  
\usepackage{graphicx} 
\usepackage{natbib}  
\usepackage{caption} 
\usepackage{algorithm}
\usepackage{algorithmic}

\usepackage{listings}
\DeclareCaptionStyle{ruled}{labelfont=normalfont,labelsep=colon,strut=off} 
\title{AAAI Press Formatting Instructions \\for Authors Using \LaTeX{} --- A Guide}
\author{
    Rada Mihalcea\textsuperscript{\rm 1}\equalcontrib, Oana Ignat\textsuperscript{\rm 2}\equalcontrib, Longju Bai\textsuperscript{\rm 1}, Angana Borah\textsuperscript{\rm 1}, Luis Chiruzzo\textsuperscript{\rm 3}, Zhijing Jin\textsuperscript{\rm 4}, \\Claude Kwizera\textsuperscript{\rm 5}, Joan Nwatu\textsuperscript{\rm 1}, Soujanya Poria\textsuperscript{\rm 6},Thamar Solorio\textsuperscript{\rm 7}  \\
}
\affiliations{
    \textsuperscript{\rm 1}University of Michigan USA, \textsuperscript{\rm 2}University of Santa Clara USA, \textsuperscript{\rm 3}Universidad de la Republica Uruguay, \\
    \textsuperscript{\rm 4}Max Plank Institute Germany, \textsuperscript{\rm 5}CMU Africa, \textsuperscript{\rm 6}SUTD Singapore, \textsuperscript{\rm 6}MBZUAI United Arab Emirates\\

}

\usepackage{bibentry}

\usepackage{comment}
    
\usepackage[dvipsnames]{xcolor}

\usepackage{url}
\usepackage{amsmath}
\usepackage{fancybox}
\usepackage{tcolorbox}

\usepackage{amsmath}
\usepackage{latexsym}
\usepackage[capitalize]{cleveref}
\usepackage{microtype}
\usepackage[T1]{fontenc}
\usepackage[utf8]{inputenc}
\usepackage{tabularx}
\usepackage{booktabs}
\usepackage{multirow}
\usepackage{xspace}
\usepackage{enumitem}
\setlist[enumerate]{itemsep=0mm}
\usepackage{fontawesome} 
\usepackage{amssymb}
\usepackage{pifont}

\makeatletter
\g@addto@macro\@floatboxreset\centering
\makeatother

\usepackage{xspace}
\graphicspath{{Figures/}}

\makeatletter
\DeclareRobustCommand\onedot{\futurelet\@let@token\@onedot}
\def\@onedot{\ifx\@let@token.\else.\null\fi\xspace}

\makeatother

\setkeys{Gin}{width=\textwidth}
\title{Why AI Is W.E.I.R.D. and Should Not Be This Way}

\title{Why AI Is WEIRD and Should Not Be This Way: \\
Towards AI For Everyone, With Everyone, By Everyone}

\begin{document}

\maketitle

\begin{abstract}
This paper presents a vision for creating AI systems that are inclusive at every stage of development, from data collection to model design and evaluation. We address key limitations in the current AI pipeline and its WEIRD\footnote{WEIRD, an acronym coined by \citep{henrich2010weirdest} to highlight the coverage limitations of many psychological studies, refers to populations that are Western, Educated, Industrialized, Rich, and Democratic. While we do not fully adopt this term for AI, as its current scope does not perfectly align with the WEIRD dimensions, we believe that today's AI has a similarly ``weird'' coverage, particularly in terms of who is involved in its development and who benefits from it. }  representation, such as lack of data diversity, biases in model performance, and narrow evaluation metrics. We also focus on the need for diverse representation among the developers of these systems, as well as incentives that are not skewed toward certain groups. We highlight opportunities to develop  AI systems that are for everyone (with diverse stakeholders in mind), with everyone (inclusive of diverse data and annotators), and by everyone (designed and developed by a globally diverse  workforce).
\end{abstract}

\section*{Introduction}

AI, and especially Large Language and Multimodal Models (LLMs and LMMs), have taken the world by storm, and yet much of the world is not represented in the data, models, and evaluations used in their development~\citep{hershcovich-etal-2022-challenges, 10.1145/3630106.3658967, nayak2024benchmarking}. This lack of representation has two major implications. First, it can lead to numerous mistakes, misconceptions, and even harms, which can propagate to the growing number of applications powered by these models, and can limit the ability of AI systems to effectively serve diverse groups and contexts. 

Second, as anthropologists have pointed out, our success as a human species is not as much due to our intelligence, as it is to our ``collective cultural brains'' that allow us to learn from one another over generations and across cultures \cite{henrich2015secret}. Cultural evolution has led to many innovations and entire bodies of knowledge 
-- a form of collective intelligence that explains our species' uniqueness and success. 

With the rapid growth of AI, we are now facing an evolution dilemma. On one side, we have our ``recipe for success'' learned over tens of thousands of years, where our collective cultural brains lead to innovation and evolution. 
On the other side, we have these very large AI models which, despite encompassing enormous bodies of information, act as a single ``super-human'' that homogenizes and erases entire bodies of cultural knowledge \cite{schwobel-etal-2023-geographical, mcveety2024digital, byrd2023truth, perez2024llms, wachter2024large,naous-etal-2024-beer}.

\begin{figure}[t]
    \centering
    \includegraphics[width=1.0\columnwidth]{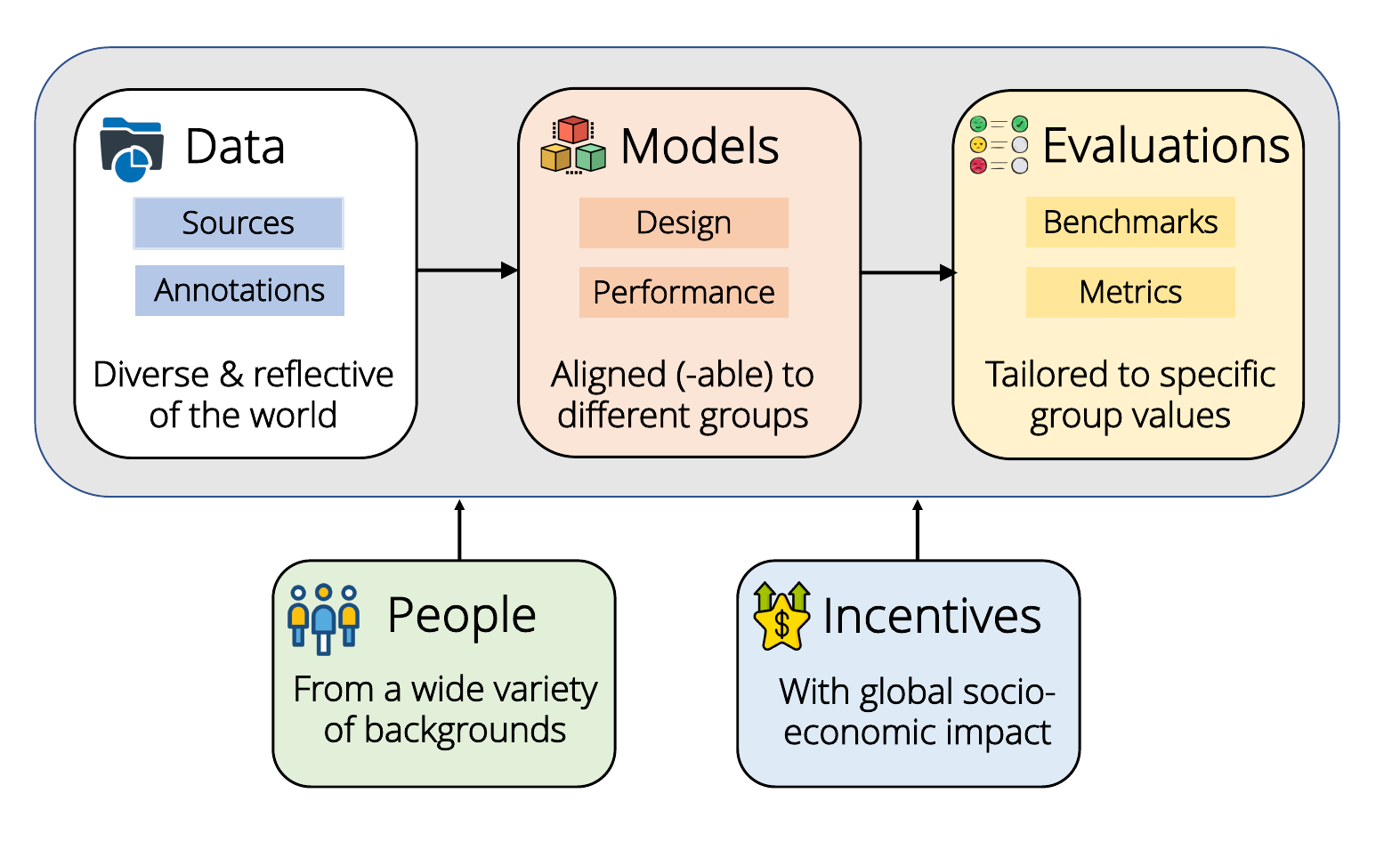}
\vskip -0.1in
    \caption{\small Desiderata and areas of research to expand the reach and impact of AI to everyone.}
\vskip -0.2in    
\label{fig:char}
\end{figure}

The goal of this paper is to present a vision for addressing limitations throughout the entire AI pipeline, including data, models, and evaluations, as well as the people driving this process and the incentives that shape its development. The aim is to ensure that these AI systems are: (1) {\bf For Everyone}: AI systems that represent everyone (\S\ref{sec:representation}), with models with even performance across groups (\S\ref{sec:modelPerformance}), inclusive evaluation metrics (\S\ref{sec:evaluationMetrics}) and culturally diverse benchmarks (\S\ref{sec:evaluationBenchmarks}), and incentives that promote inclusive AI and balance profits and social impact (\S\ref{sec:incentives}); (2) {\bf With Everyone}: relying on diverse data sources (\S\ref{sec:DataSources}), diverse annotators and inclusive annotation standards (\S\ref{sec:DataAnnotations}); (3) {\bf By Everyone}:  model designs that are unbiased (\S\ref{sec:modelDesign}), built by a diverse group of developers and leading to applications grounded in real-life (\S\ref{sec:People}).

We assembled a team of authors with broad expertise in the field of AI, who through their current or native countries, bring insights from twelve different cultures (China, Germany, India, Mexico, Nigeria, Romania, Rwanda, Singapore, Switzerland, United Arab Emirates, United States, Uruguay).

\section{Representation in AI}\label{sec:representation}

Just as it is critical to see ourselves represented in our communities, we also strive for representation in the AI systems we use. Yet, current models often fail to grasp the characteristics of different cultures, either because they lack this cultural knowledge or because they fail to recognize when and where to apply it. To further complicate this, the definition of ``culture'' encompasses many aspects, along multiple semantic and demographic axes \cite{thompson2020cultural,adilazuarda2024towards}, including among others social norms (e.g., gift giving), beliefs and habits (e.g., daily routines), artistic taste (e.g., traditional music), or subjective perceptions (e.g., emotional connotations). 
Further, the narratives produced by these models are frequently seen through an outsider's lens, missing the essential culture-centric perspectives that give depth and authenticity, as illustrated in the vignette in \Cref{fig:vignetteOutsider}. 

\begin{figure}[h]
\vskip -0.1in
\begin{tcolorbox}[width=.95\linewidth, colframe=black, colback=yellow!15, boxsep=0mm, boxrule=1.5pt, arc=3mm]
\it \small
Petru is an 11-year-old boy living in his home country of Romania, at the age when he looks for role models to emulate. He asks an AI system for examples of male role models, and the reply comes back confidently: Nicolae Ceau\c{s}escu, with the justification that ``he played a significant role in the Romanian history [...] and his regime had a lasting impact.'' This answer misleads Petru to believe that a dictator who was one of the darkest figures in Romanian history should be a model to follow. 
\end{tcolorbox}
\vskip -0.1in
\caption{\small Outsider perspectives on the history and culture of groups not represented in AI models often conflict with the insider perspectives and can be misleading.}
\vskip -0.2in
\label{fig:vignetteOutsider}
\end{figure}

\paragraph{Lack of Cultural Knowledge.}

The majority of resources used in the development of AI models often consider only general knowledge and disregard the cultural aspect \cite{shen2024understanding}. This knowledge permeates the entire AI pipeline, from data distribution and annotation (\S \ref{sec:data}), to model development and alignment (\S \ref{sec:models}), to evaluation metrics and benchmarks (\S \ref{sec:evaluation}). Lack of representation can inevitably introduce biases, which can stem from how the data collection is framed by the dataset developers \cite{Parmar2022DontBT}, from the annotators' belief \cite{Sap2021AnnotatorsWA} or the background of the contributors \cite{Aguinis2020MTurkRR}, from the strategies used to fine-tune or align AI models ~\cite{ouyang2022training, zhang2023instruction}, or from the benchmarks used to assess model performance \cite{shen2024understanding}.  

\noindent \underline{Opportunities.} Recent work included efforts to expand the cultural knowledge encompassed by AI systems along dimensions of time \citet{Shwartz2022GoodNA}; food-related customs \cite{Palta2023FORKAB}; factual knowledge  \cite{Yin2022GeoMLAMAGC,Nguyen2022ExtractingCC,keleg-magdy-2023-dlama,romero2024cvqa}; or cultural norms \cite{fung-etal-2023-normsage}. Scaling these diversification efforts to include the numerous cultures worldwide and their many cultural dimensions is an ongoing challenge, particularly for those cultures with limited online representation. This may require close engagement with cultural experts or anthropologists who have an insider view into the culture of a group and understand its heritage~\citep{10.1145/3534970}. Additionally, it may necessitate developing tailored strategies to engage with cultural groups ``where they are,'' acknowledging that not everyone has an online presence, and that much of this cultural knowledge floats openly in forms that may not be easily captured in digital form. Finally, people often affiliate with more than one culture, which requires solutions for cultural compositionality~\citep{welch-etal-2020-compositional} or pluralistic alignment~\citep{sorensen2024roadmap}.

\section{\includegraphics[width=0.045\textwidth]{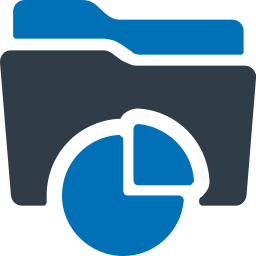} Data}\label{sec:data}

Data is an essential component of the AI pipeline, used to train or fine-tune models but also to evaluate their performance.  As illustrated in the vignette in \Cref{fig:vignette_annotationsponge}, a lack of diverse representation in AI datasets can have a negative impact. In the process of creating datasets, two aspects play a major role: the source of the data and its annotation. 

\begin{figure}[h]
\begin{tcolorbox}[width=.95\linewidth, colframe=black, colback=yellow!15, boxsep=0mm, boxrule=1.5pt, arc=3mm]
\it \small
Chinasa wants her small business to cater to a global audience, so she creates a website to sell African bath sponges and uploads several photos of her products to the site. Seeing that sales do not take off, she decides to do a simple image search for the word ``sponge'' on an AI-powered web search engine, just to find that out of a hundred image results, there is only one image that looks like the kind of sponge she sells. 
\end{tcolorbox}
\vskip -0.1in
\caption{\small Models trained on non-inclusive datasets hinder the representation of stakeholders in mainstream media.
}
\vskip -0.2in
\label{fig:vignette_annotationsponge}
\end{figure}

\subsection{Data Sources}\label{sec:DataSources}

\paragraph{Lack of Data Transparency.}
AI models, and especially LLMs and LMMs, rely on increasingly large collections of data. However, because of the current fast speed of research with minimal or non-existent regulations and standards, we have witnessed a crisis in data transparency~\cite{longpre2023data}, often associated with unwanted societal biases and unexpected behaviors ~\citep{buolamwini2018gender, gebru2020race}. In response to these challenges, several strategies have emerged to enhance data documentation in AI. \citet{gebru2021datasheets} introduced the concept of datasheets for datasets, which was followed by the development of data statements and data cards for NLP data~\cite{bender-friedman-2018-data, holland2020dataset, Pushkarna2022DataCP, longpre2023data}. 

\noindent \underline{Opportunities.} 
Future efforts must prioritize the development of transparent, responsible, and data-centric AI models. 
Data licenses have the potential to promote more responsible, inclusive, and transparent machine-learning practices, but work still needs to be done to understand how to define and interpret license terms for data usage, and how to adapt them to AI data and models \citet{longpre2023data}. AI researchers need to collaborate with policymakers and legal experts to develop tools that empower and educate users and creators about dataset documentation from various perspectives, including provenance, identifiers, and characteristics~\cite{longpre2023data}. Inspiration can be drawn from the database community~\cite{bhardwaj2014datahub} and more regulated fields such as medicine~\cite{Moherc869}. Additionally, innovative ideas from the NLP community, such as leveraging LLMs to document data and models, can further complement these efforts~\cite{liu-etal-2024-automatic}.

\paragraph{Lack of Data Diversity.}
The predominant methods for data collection involve scraping vast amounts of text and image data from the Web~\cite{villalobos2022will}. While effective for Western communities, this approach excludes minority groups with limited or no internet access, representing 37\% of the global population~\cite{nwatu2023bridging}.
Moreover, there are many communities for which  producing written data is a major challenge~\cite{wiechetek2024ethical}.
Further, these uneven distribution and biases in data are perpetuated by the current practice of self-supervised learning where the data is labeled by AI models trained from a Western perspective~\cite{oquab2023dinov2, Ramaswamy2023BeyondWC}.

\noindent \underline{Opportunities.} We need to reevaluate our current data collection practices, and collect data that covers a wide range of perspectives across demographic and cultural dimensions. Recent work has shown that even small amounts of diverse data can improve performance \cite{ramaswamy2024geode}. 
Efforts to improve representation in AI and reduce the need for large datasets include active learning~\citep{hady2013semi}, domain adaptation~\citep{kalluri2023geonet, wang2023overwriting}, similarity-based~\cite{ignat2024annotations} and grammar-based~\cite{lucas2024grammar} strategies for data augmentation.
Additionally, the development of diverse datasets should involve people from various demographics to guide data collection and annotation, foster collaboration, and empower stakeholders~\citep{nwatu2023bridging}.

\subsection{Data Annotations} \label{sec:DataAnnotations}

\paragraph{Lack of Inclusive Annotation Standards.}
Most annotation standards fail to account for the diversity and subjectivity inherent in global data~\citep{nwatu2023bridging}. 
Most benchmarks adhere to popular and conventional annotation systems such as ImageNet \cite{Deng2009}, which however are not a definitive standard~\cite{beyer2020we, fang2023, shankar2020} and have not been designed with cultural representation in mind.  As a result, models trained on these benchmarks tend to overfit to an incomplete  'gold standard' that does not accurately represent the world as we see it \cite{shankar2020, mayfield-etal-2019-equity}, and can lead to models that do not  generalize well to real-world tasks and out-of-distribution data \cite{fang2023, taori2020measuring}. 

\noindent \underline{Opportunities.}
A significant challenge is defining and promoting standards that address common annotation issues encountered when collecting and annotating globally diverse data.
Several studies \cite{beyer2020we, yun2021re, shankar2020, faghri2023reinforce} have identified issues such as label noise, errors, ambiguity, and restrictiveness in current datasets, proposing various methodologies to improve these labels.
Solutions include generating robust labels using strong pre-trained models, integrating human feedback,  incorporating multiple labels per image, or explicitly accounting for the diversity of annotators \cite{deng-etal-2023-annotate}. Flexible data annotation structures that use e.g., function-based labeling could also offer a strategy for improving representation in AI datasets~\citep{nwatu2023bridging}.

\paragraph{Lack of Diverse Annotators and Curators.}
An demographic report  of the most popular annotation platform (Amazon Mechanical Turk) \cite{difallah2018demographics} found that the majority (75\%) of annotators are from the United States and 16\% are from India. This is concerning, since the demographic background was found to significantly influence the people's ratings and performance  \citet{pei2023annotator}. 
Similar statistics are observed in many existing AI datasets \cite{kirk2024prism}.

\noindent \underline{Opportunities.}
New annotation frameworks may be required for inclusive data annotations that address the unique aspects of a problem. 
The MaRVL dataset \cite{liu-etal-2021-visually} and CVQA dataset \cite{romero2024cvqa} are good examples of projects that prioritize inclusive data annotation by involving native speakers in selecting concepts, questions, and images.
Additionally, it is important to maintain transparency on the annotation process \cite{gebru2021datasheets}, and share 
aggregate statistics for the distribution of workers by region and demographics \cite{kirillov2023segment}.
This information, combined with methods that identify demographic blind spots across datasets \cite{dominguez2023dsap} can help users make informed decisions about which datasets to use for their specific applications.

\vspace{-0.1in}

\section{\includegraphics[width=0.05\textwidth]{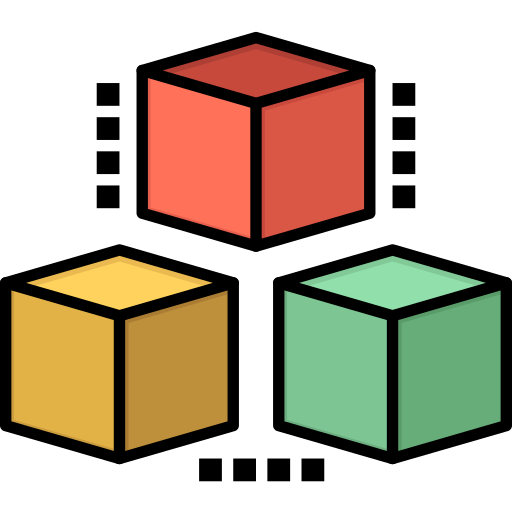} Models}\label{sec:models}

AI models are the engines that drive the capabilities of modern AI systems, enabling them to recognize patterns, make predictions, and generate content. Their effectiveness is heavily influenced by their architecture, and as illustrated in Figure \ref{fig:vignetteModel}, their performance can be uneven and limited to specific contexts and settings. 

\begin{figure}[h]
\begin{tcolorbox}[width=.95\linewidth, colframe=black, colback=yellow!15, boxsep=0mm, boxrule=1.5pt, arc=3mm]
\it \small
Maya is a high school administrator in a multicultural urban area in Canada. She decides to use a new AI-driven educational tool  to help her personalize the learning experiences for her students. She soon notices that the tool performs poorly when students input text in the local French dialect, often misunderstanding the context or giving the wrong output, an issue not faced by her English-speaking students.  

\end{tcolorbox}
\vskip -0.1in
\caption{\small Uneven model performance for different languages leads to incorrect output and can favor those speaking the languages for which the model performs better.}
\vskip -0.2in
\label{fig:vignetteModel}
\end{figure}

\subsection{Model Design} \label{sec:modelDesign}
\paragraph{Models Exhibit Biases in  Pre-training and Alignment.}
AI models acquire their {\it knowledge} primarily during the pre-training phase; later, the specification for how to interact with this knowledge is provided by the process of \textit{alignment}. The encoded knowledge is mainly determined by the dataset used in the model pre-training and thus a crucial factor for potential bias in the model; we refer to this bias as \textbf{knowledge bias}. The alignment process allows us to interact with the encoded knowledge by acting as a knowledge retriever, and can shift the model toward certain preferences; we refer to this as \textbf{alignment bias}. However, several recent investigations have demonstrated that AI models consistently excel in US-specific queries while struggling with underrepresented cultures \cite{shen2024understanding,masoud2024culturalalignmentlargelanguage}. 

\noindent \underline{Opportunities.}
To mitigate knowledge bias, the primary approach is to pre-train or fine-tune LLMs on culture-specific data \cite{li2024culturellmincorporatingculturaldifferences}, or to rely on  prompt engineering \cite{wang2024countriescelebratethanksgivingcultural,kovač2023largelanguagemodelssuperpositions,rao2023ethicalreasoningmoralalignment}. 
These approaches raise the important question of how to effectively gather preference data that is reflective of diverse cultures. One strategy is to manually collect preference data based on Hofstede's Cultural Alignment Test~\cite{masoud2024culturalalignmentlargelanguage}. Alternatively, LLMs could engage in conversations with individuals from various cultures to generate preference data, for instance by using multi-agent settings ~\cite{li2024cultureparkboostingcrossculturalunderstanding}. 
An additional challenge is the integration of diverse preferences into a single model using strategies for pluralistic alignment \cite{sorensen2024roadmap}. or meta-reward models. 

\subsection{Model Performance} \label{sec:modelPerformance}

\paragraph{Poor Model Generalization.}

AI models often struggle to maintain performance when encountering diverse and dynamic real-world settings \cite{malik2024objectcompose, gustafson2023pinpointing}. These performance issues are further complicated by differences in geographical locations and income levels, leading to disparities in model performance across different regions and social strata \cite{rojas2022the, ramaswamy2023geode, nwatu2023bridging}. Similarly, we see biased outcomes and reduced usability across diverse linguistic communities, affecting fairness and inclusivity in AI applications \cite{ziems-etal-2022-value, blodgett-etal-2018-twitter, xiao-etal-2023-task, malmasi-etal-2016-discriminating, zhou-etal-2021-challenges}, or potentially generating responses that are inappropriate or offensive in different cultural contexts \cite{peterson2023measure, tay-etal-2020-rather, santurkar2023opinionslanguagemodelsreflect, anonymous2024from, moore2024largelanguagemodelsconsistent}.

\noindent \underline{Opportunities.} Improving the model generalization across diverse contexts can involve active learning or other data-driven strategies that can help supplement data for underrepresented groups \cite{ignat2024annotations}. Alternatively, specialized benchmarks (e.g., VALUE \cite{ziems-etal-2022-value} or DADA \citet{liu-etal-2023-dada}) can handle the languages spoken by diverse groups more effectively \cite{xiao-etal-2023-task,  sun-etal-2023-dialect, hofmann2024dialectprejudicepredictsai, faisal-etal-2024-dialectbench}.
Another direction can consider benchmarks and  training processes that account for multilingual and multicultural nuances, following  initiatives such as the ETHICS dataset ~\cite{hendrycks2021aligning}, or frameworks  to assess and enhance  AI alignment with diverse human values \cite{peterson2023measure, choenni-etal-2024-echoes}.

\paragraph{Security Vulnerabilities and Propagation of Harmful Stereotypes.}
AI models are increasingly vulnerable to security breaches and the propagation of harmful stereotypes, and \textit{jailbreaking} attacks pose a significant threat by circumventing the safety mechanisms designed to prevent models from generating unethical, harmful, or dangerous content  \cite{ouyang2022training, rafailov2023direct}. 
This is paricularly true for low-resource languages, making underrepresented groups especially vulnerable \cite{yonglow23}. Similarly, AI models are prone to perpetuating the stereotypes embedded in their training data, reinforcing societal prejudices related to race, gender, and ethnicity, among others \cite{ferrara_fairness_2023, hofmann2024dialect}.

\noindent \underline{Opportunities.}
Addressing these intertwined challenges requires more robust alignment and defense strategies, and specialized benchmarks \cite{shu2024attackeval,mazeika2024harmbench,luo2024jailbreakv,liu2024mmsafetybench} that can lead to a deeper understanding of the underlying mechanisms of harmful outputs and security breaches. 
Mechanistic interpretability for alignment algorithms \cite{lee2024mechanistic} also offers promising avenues for controlling model responses and mitigating the success of both jailbreak attacks and the propagation of harmful stereotypes \cite{arditi2024refusal, ball2024understanding}.
Combating the propagation of harmful stereotypes will require innovations in both the training \cite{li-etal-2023-prompt,kumar-etal-2023-parameter} and post-training \cite{pmlr-v162-ravfogel22a,cheng2021fairfil} phases of model development. Additionally, direct adjustments to model architectures, such as integrating awareness of harmful outputs into operational frameworks or adjusting the attention to different social groups \cite{gaci-etal-2022-debiasing,kim-etal-2024-able-agency} can also help ensure the fairness across different social groups.

\section{\includegraphics[width=0.045\textwidth]{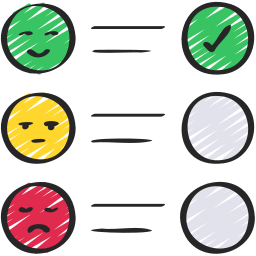} Evaluation}\label{sec:evaluation}

Evaluating AI systems is essential for ensuring their accuracy and reliability across various applications. This process typically focuses on metrics and benchmarks, which are used to measure performance and identify any biases or limitations.

\begin{figure}[h]
\vskip -0.1in 
\begin{tcolorbox}[width=.95\linewidth, colframe=black, colback=yellow!15, boxsep=0mm, boxrule=1.5pt, arc=3mm]
\it \small
Aarav, a developer passionate about empowering Indian children through education, wants to deploy an AI-powered education tool in several Indian villages. He soon realizes that the tool was evaluated using metrics designed for Western learning styles, focusing on individual achievement and competition. When deployed in India's collectivist society that values group collaboration and shared success, the tool fails to resonate with students and produces misleading performance results. 
\end{tcolorbox}
\label{fig:vignette_eval}
\vskip -0.1in
\caption{\small A misalignment between evaluation metrics and cultural values can lead to misleading estimation of a tool's effectiveness.}
\vskip -0.2in
\end{figure}

\subsection{Evaluation Metrics}\label{sec:evaluationMetrics}
\paragraph{Lack of Inclusive Metrics.} 
While several of the metrics used to evaluate AI models may be considered generic, such as accuracy or F1 scores, there are also many metrics that only reflect the reality of specific populations. For example, reading comprehension metrics may assume familiarity with Western literary references \cite{steffensen1979cross, Kolisko_Anderson_2024}. Similarly, bias detection metrics, such as the Word Embedding Association Test (WEAT) \cite{caliskan2017semantics} and Sentence Association Embedding Test (SEAT) \cite{may-etal-2019-measuring} rely on Western-centric norms \cite{greenwald1998measuring}. 
 Additionally, the human evaluations sometime used to assess system performance --through crowdsourcing, expert evaluations, or user studies -- while offering a more nuanced understanding of model performance, they also heavily depend on and can be biased by the background of the evaluators \cite{song2013bleu, reiter2018structured}. 
 
\noindent \underline{Opportunities.} 
To ensure a comprehensive and fair assessment, generic metrics such as accuracy and F1 scores should be coupled with diverse and inclusive datasets and metrics that assess performance across all majority and minority classes. 
For metrics that are inspired by the reality of specific groups, it is critical to adjust them to reflect the needs and values of those directly impacted. Combining human evaluations with automatic metrics can enhance the reliability of assessments \cite{van2021human, schuff2023human}, particularly when the human evaluators represent a diverse range of backgrounds.  
Additionally, fairness metrics such as demographic parity \cite{10.1145/2090236.2090255}, equalized odds, and statistical parity \cite{hardt2016equalityopportunitysupervisedlearning} should be used alongside traditional performance metrics to identify and address biases in AI models. 

\subsection{Evaluation Benchmarks}\label{sec:evaluationBenchmarks}
\paragraph{Lack of Culturally Diverse Benchmarks.} 
Most benchmarks are heavily biased towards English-speaking and Western cultures, often focusing on limited datasets that do not consider the visual or language diversity across cultures. 
Furthermore, numerous languages are frequently overlooked in benchmark construction, and this is especially true for languages that do not have a written form. 
Particularly challenging are sign languages \cite{baker2015sign}, and benchmarks based solely on spoken languages \cite{pine2017language}. 

\noindent \underline{Opportunities.} 
Culturally and linguistically diverse benchmarks are crucial for developing models that can be applied globally. This involves diversifying data sources (\S \ref{sec:DataSources}), and ensuring the annotators are familiar with the cultural or linguistic context (\S \ref{sec:DataAnnotations}). Collaboration among researchers, linguists, and cultural experts is also key to developing benchmarks that are truly representative \cite{oheigeartaigh2020overcoming}. Multimodal benchmarks are particularly promising as they can provide valuable context for models. 
In languages that rely solely on speech, preserving oral literature often entails documentation and collaboration with native speakers to annotate images and videos \cite{bird2010scalable, leedom2014language}. 
For sign languages, ongoing efforts are being made to develop writing systems like HamNoSys \cite{hanke2004hamnosys} or SignWriting \cite{sutton2010signwriting}, although there is currently no universally accepted standard. 

\section{\includegraphics[width=0.045\textwidth]{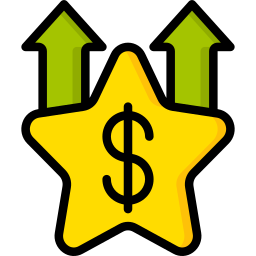} Incentives}\label{sec:incentives}\label{sec:Incentives}

The incentives that drive the prioritization of certain cultures and languages in current AI technology development can vary widely \cite{joshi2020state,bommasani2021opportunities,bird2020decolonising,rogers2021changing}, and are often connected to the source of funding behind the AI initiatives: economic drives, government support, or philanthropic initiatives. These, in turn, can influence individual decisions, as illustrated in Figure \ref{fig:vignetteincentives}. 

\begin{figure}[h]
\vskip -0.1in
\begin{tcolorbox}[width=.95\linewidth, colframe=black, colback=yellow!15, boxsep=0mm, boxrule=1.5pt, arc=3mm]
\it \small
Maria is the CEO of a small tech startup in Romania, with limited staff and research resources. She wants to build an AI assistant that can help people with their daily chores. However, as she delves deeper, she quickly faces a hard problem:
Should she target a high purchasing power demographic like the United States and Western Europe? Or should she focus on a market where cultural values align more closely with her vision, such as Romania and Eastern Europe, even if the financial returns are less certain? 
\end{tcolorbox}
\vskip -0.1in
\caption{\small Selecting the right market for AI products requires balancing cultural values and financial returns in a high-stakes decision.}
\vskip -0.1in
\label{fig:vignetteincentives}
\end{figure}

\paragraph{Economic Drives Prioritize Rich Countries and Major Languages.}
One of the largest root causes for AI development lies in its economic value, both immediate and potential \cite{furman2019ai,chui2023economic}. There is generally a lower perceived profit rate for developing AI applications for smaller groups or those from lower social and economic levels \cite{joshi2020state,blasi2022systematic}. 
However, this raises the question of whether the perceived profit matches the actual potential profit, as people may underestimate the unique opportunities within smaller communities \cite{bird2020decolonising,nekoto2020participatory}. Moreover, it introduces the efficiency versus fairness dilemma, as pursuing market-driven profits can exacerbate social inequalities and stability, often referred to as the ``rich get richer'' effect \cite{hovy2016social,bender2021dangers}.

\noindent \underline{Opportunities.}
To mitigate the concentration of economic investments in highly profitable areas, we need strategies to encourage companies to balance fairness with profits \cite{crawford2021atlas,bender2021dangers}. Governments and philanthropic organizations can create incentives to address fairness by bridging the investment gap in less profitable areas, and ensuring that the perceived economic value of an investment is not the only drive behind AI development decisions.

\paragraph{Inconsistent Government Support for Inclusive AI.}
Governments often support technologies with immediate or potential socio-economic impact, as well as those essential for national defense \cite{wolff2012rising,weiss2014america,fleming2019government}.
While some support, e.g., for fundamental science research, can promote the development of AI technologies adaptable to underrepresented languages and communities, this support is not always consistent \cite{shibayama2011distribution,NSFUnderrepresented2024}. Some governments lack the resources for such investments, requiring justifications for their expenditure \cite{bird2020decolonising,nekoto2020participatory}. 
Furthermore, prioritizing technologies that maintain a competitive national advantage can create a dilemma, as nations that invest in  ``good causes'' to help underrepresented communities, domestic or international, may fall behind those that do not \cite{okun2010equality,berg2011equality}.

\noindent \underline{Opportunities.}
It is essential to improve decision-making systems to justify expenditures on fairness and support for underprivileged groups. Research demonstrating that such investments also enhance the quality of life for most citizens can bolster support. Moreover, these decisions should take into account sensitive international contexts and strive to balance maintaining national strength with ensuring equitable AI benefits for diverse populations.

\paragraph{Insufficient Philanthropic Initiatives for Promoting Inclusive AI.}
Philanthropic organizations play a vital role in counterbalancing the market-driven allocation of resources by supporting needed areas where funding does not naturally flow  \cite{brest2018money,brass2018ngos}.
Although philanthropies are well-positioned to address resource distribution issues, advocating for increased support for underrepresented AI can be challenging \cite{jammulamadaka2010ngo}, as these philanthropic organizations must weigh investments to help various other needs \cite{givewell2022metrics}. 

\noindent \underline{Opportunities.} Encouraging more philanthropic support of AI development requires carefully thought-out justifications regarding the long-term value of research and development \cite{pennings1997option,pisano2012creating}. Helping disadvantaged groups can be approached by either providing immediate aid, or by enhancing their skills and education through technology, which leads to increased productivity \cite{kuhn2020reducing,saiz2018tackling}. Therefore, advocating for AI technology investment should emphasize its potential to unlock significant long-term benefits for these groups.

\section{\includegraphics[width=0.045\textwidth]{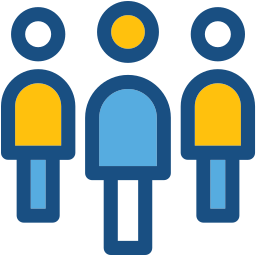} People} \label{sec:People}

Although AI has the potential for global impact, research and development are dominated by a few countries~\citep{kleinberg2021algorithmic, hershcovich-etal-2022-challenges, alkhamissi2024investigating}. Our limited experiences can blind us to biases against the very communities we aim to help, even with the best intentions, as shown in \Cref{fig:vignettePeople}.

\begin{figure}[h]
\begin{tcolorbox}[width=.95\linewidth, colframe=black, colback=yellow!15, boxsep=0mm, boxrule=1.5pt, arc=3mm]
\it \small 
A healthcare AI company launches an app to streamline patient diagnoses in a rural Nigerian community without accounting for local medical practices and resources. The app misinterprets common regional symptoms, like skin rashes from heat and sun exposure, as severe conditions, resulting in unnecessary treatments. By not involving local healthcare providers and patients in its development, the app ultimately causes more harm than good.
\end{tcolorbox}
\vskip -0.1in
\caption{\small People affected by AI systems are often not involved in their development, often leading to applications that fail to addresses real problems or even cause harm.}
\vskip -0.2in
\label{fig:vignettePeople}
\end{figure}

\paragraph{Lack of Global Diverse Representation and Agency in AI Research.}

If AI systems reinforce dominant cultures, whether implicitly or explicitly, they might lead to a cycle of cultural homogeneity~\citep{schramowski2022large, vaccino2023exploring}.
To ensure that the systems and resulting applications reflect people's authentic culture, representatives from these groups should also participate in the design, data construction, and model development process in a  participatory approach \cite{bondi2021envisioning}. Having a diverse group of developers is essential to  foster innovation that addresses the needs and values of different cultures \cite{page2010diversity}. 
Reciprocity and mutual learning are central to successful research engagements~\cite {brereton2014beyond, taylor2019relational, st2022reimagining}.

\noindent \underline{Opportunities.}
We need to reevaluate the power dynamics between technologists and community members and establish equal research partnerships with the community~\citep{bird2024centering, mignolo2012local}. 
This approach follows a decolonizing practice that respects the sovereignty of local communities and prioritizes their input~\citep{bird2020decolonising}. 
Additionally, several workshops and events have begun to explore how to empower stakeholders in the development and deployment of technology~\citep{vaccaro2019contestability, givens2020centering} and how to help researchers and practitioners consider when not to build systems at all~\citep{barocas2020not}.
Other examples of offering mentorship are open-source and free initiatives such as Masakhane,\footnote{{https://masakhane.io}} Black in AI,\footnote{{https://blackinai.github.io}}, 
AmericasNLP,\footnote{{https://turing.iimas.unam.mx/americasnlp}} and the ACL Mentorship,\footnote{{https://mentorship.aclweb.org}} just to mention a few.

\paragraph{Lack of Practical Applications and Real-Life Grounding in AI Research.}

Language is inherently situated. 
However, many AI systems and AI research fail to clearly articulate what problems they tackle.
For instance, a survey of 146 papers on bias in NLP systems  shows that their motivations are often vague, inconsistent, and lack normative reasoning \cite{blodgett-etal-2020-language}. 
Furthermore, different social groups, especially those at the intersections of multiple axes of oppression, have different lived experiences due to their different social positions~\citep{sassaman2020creating, field-etal-2021-survey}.

\noindent \underline{Opportunities.} 
To ensure our AI models effectively serve the communities they are intended to help, it is crucial to first understand their specific needs and how they plan to use the technology. This requires focusing on the lived experiences of those directly impacted by these systems~\citep{blodgett-etal-2020-language}. 
Cross-disciplinary collaboration is particularly important for real-world impact. For example, sociolinguists and anthropologists have studied how language varieties are perceived—whether as standard, correct, or uneducated~\citep{reaser2018language, roche2019articulating, craft2020language}. This research reveals that beliefs about language often mirror deeper beliefs about the speakers themselves~\cite{rosa2017unsettling}. Understanding the role of language in maintaining social hierarchies is vital for improving bias analysis in NLP systems and addressing how racial ideologies both shape and are shaped by technology~\citep{ruha2024}.

\section{Conclusion}

One of the key concerns about the rapid integration of AI technologies into everyday life is the risk of widening the socio-economic gap by disproportionately benefiting certain groups while marginalizing others, as well as the potentially negative impact it can have on individuals and society. 
On an individual level, it affects people's perception of themselves and others, influencing their interactions and the opportunities they have access to. 
On a societal level, the widespread use of non-inclusive AI systems can shape cultural norms and social structures and support discriminatory narratives that hinder efforts toward equality and inclusivity.

The vision we lay out in this paper highlights opportunities to develop AI for everyone, with everyone, by everyone. Our key recommendations focus on improving inclusivity along the entire AI pipeline, specifically targeting the five main areas we addressed in this paper: {\bf Data}: Prioritize the development of transparent, responsible, and data-centric AI models; Reevaluate current data collection practices to ensure coverage of diverse perspectives across demographic and cultural dimensions; Define inclusive annotation standards to improve the representation in training data; Adopt a participatory approach to AI, engaging community members early in the research process. {\bf Models}: Architect model designs that minimize knowledge and alignment bias;  Improve model generalization to ensure performance across diverse groups and contexts; Build models that are robust and not vulnerable when used by underrepresented groups. {\bf Evaluations}: Create inclusive  metrics that assess performance across both majority and minority groups, and accurately reflect the reality of the target users;  Develop culturally and linguistically diverse evaluation benchmarks. {\bf Incentives: } Devise economic strategies to encourage companies to balance fairness with profits;  Shape government agendas that promote both national strength and equitable AI benefits for all;  Encourage philanthropic support of long-term AI research agendas focused on inclusivity. {\bf People.} Foster the growth of a diverse AI workforce; Establish equal research partnerships with communities;  Focus on the lived experiences of those directly impacted by these systems.

Each of these recommendations requires concrete steps to ensure the AI systems are equitable, robust, and representative of all populations. By addressing these core areas, we can advance towards  AI systems that serve everyone, are built with input from a wide range of perspectives, and reflect the contributions of a diverse group of  stakeholders.

\bibliography{main.bib,custom.bib}

\end{document}